\documentclass[pre,twocolumn,superscriptaddress,showpacs,floatfix]{revtex4-1}
\usepackage{graphicx}
\usepackage{amssymb,amsmath,amsbsy,amsfonts,bm}
\usepackage{epsfig}
\usepackage[percent]{overpic}
\usepackage{color}

\begin{document}
\title{Steady-state rheology and structure of soft hybrid mixtures of liquid crystals and magnetic nanoparticles}

\author{Gaurav P. Shrivastav}
%\footnote{Present address: }}
\email{gaurav.shrivastav@tuwien.ac.at}
\affiliation{Institut f\"ur Theoretische Physik, Technische Universit\"at 
Wien, Wiedner Hauptstr. 8-10/136, 1040 Vienna, Austria.}

\author{Nima H. Siboni}
\email{hamidisiboni@tu-berlin.de}
\affiliation{Institut f\"ur Theoretische Physik, Technische Universit\"at 
Berlin, Hardenberg Str. 36, 10623 Berlin, Germany.}

\author{Sabine H. L. Klapp}
\email{klapp@physik.tu-berlin.de}
\affiliation{Institut f\"ur Theoretische Physik, Technische Universit\"at 
Berlin, Hardenberg Str. 36, 10623 Berlin, Germany.}

\begin{abstract}
Using non-equilibrium molecular dynamics simulations, we study the rheology of a model hybrid mixture of liquid crystals (LCs) and dipolar soft spheres (DSS) representing magnetic nanoparticles. The bulk isotropic LC-DSS mixture is sheared with different shear rates using Lees-Edwards periodic boundary conditions. The steady-state rheological properties and the effect of the shear on the microstructure of the mixture are studied for different strengths of the dipolar coupling, $\lambda$, among the DSS. We find that at large shear rates, the mixture shows a shear-thinning behavior for all considered values of $\lambda$. At low and intermediate values of $\lambda$, a crossover from Newtonian to non-Newtonian behavior is observed as the rate of applied shear is increased. In contrast, for large values of $\lambda$, such a crossover is not observed within the range of shear rates considered. Also, the extent of the non-Newtonian regime increases as $\lambda$ is increased. These features can be understood via the shear-induced changes of the microstructure. In particular, the LCs display a shear-induced isotropic-to-nematic transition at large shear rates with a shear-rate dependent degree of nematic ordering. The DSS show a shear-induced nematic ordering only for large values of $\lambda$, where the particles self-assemble into chains. Moreover, at large $\lambda$ and low shear rates, our simulations indicate that the DSS form ferromagnetic domains.
\end{abstract}
\maketitle
%%%MAIN TEXT%%%%
\section{Introduction}
In recent years, composites of liquid crystals (LCs) and magnetic nanoparticles (MNPs) have been established as an important new class of soft "hybrid" materials.
An attractive feature of these systems, which have been originally proposed by Brochard and de Gennes \cite{gb70}, is that their structural and material properties can be tuned by external fields, such as magnetic, electric and surface fields. This makes them interesting not only from a fundamental perspective, e.g., in the context of spontaneous ferromagnetism \cite{mldc13} and magnetic field-induced nematic order \cite{mespkk16}, but also for technical and medical applications \cite{lms09,sn05,lals16}. For these reasons, mixtures of LCs and MNPs have been extensively studied in experiments \cite{ns86,ca83,ktkz08,bbc98,pbb11,bnr11,kbd13,mldc13} and, more recently, also in particle-based computer simulations \cite{plk15,pk15,pk16}. Most of these studies have been devoted to systems in thermal equilibrium.

One feature which is particularly accessible for computer simulations is the self-assembly of the MNPs into clusters and chains due to magnetic dipole-dipole interactions, typically measured by the coupling parameter $\lambda$. This unique feature of the MNPs, and its interplay with the LC host matrix, was investigated in detail in several computer simulation studies of Gay-Berne (GB) and dipolar soft spheres (DSS), i.e., spheres with a permanent point dipole moment \cite{plk15,pk15,pk16,gk19}. All of these studies were concerned with equilibrium systems at relatively low densities of MNPs. When the matrix is globally isotropic, the MNPs assemble into randomly distributed chains whose length depends on $\lambda$. Interestingly, if these chains are aligned by an external magnetic field, they can induce some degree of nematic ordering 
in the LC matrix, a feature confirmed by experiments \cite{mespkk16}. In turn, when the LC matrix undergoes a spontaneous isotropic-nematic (I-N) phase transition, it provides an anisotropic environment already in the absence of a field. The MNP chains then align along the LC director \cite{pk15}. Moreover, the nematic matrix modifies the equilibrium translational dynamics of the MNPs; they display anomalous diffusion at intermediate timescales, with the extent of the anomalous regime depending again on $\lambda$ \cite{gk19}.

While the equilibrium structure of LC-MNP mixtures is important and quite intriguing, many applications of such soft hybrid systems, as well as some experiments, actually involve nonequilibrium conditions, particularly shear flow. Taking this as a motivation, we here present results from nonequilibrium Molecular Dynamics (MD) simulations for model mixtures composed of GB and DSS particles in planar Couette shear flow, characterized by the shear rate $\dot\gamma$.
We investigate both, structural and rheological properties. To unravel the impact of shear, we use the same model potentials and parameters as in earlier simulation studies \cite{plk15,pk15,gk19}. Special interest is devoted to the role of the magnetic coupling parameter $\lambda$, which drives the chain formation and has already turned out to be crucial for the equilibrium structure. We thus expect that $\lambda$ will also crucially affect the rheology (e.g., the shear stress) of the mixtures, which may provide a path to control the flow properties of these hybrid materials even without an external magnetic field.

As a background for the shear-induced behavior of the LC-MNP mixtures we note that already pure LCs show pronounced nonlinear behavior under shear. In particular, one
observes shear thinning behavior at large $\dot\gamma$ \cite{l99,w81}, i.e., the apparent viscosity (steady-state shear stress over shear rate) decreases with increasing $\dot\gamma$. 
This shear thinning can be related to the occurrence of a shear-induced I-N transition \cite{rhw08,rhjp08,gs05}, that is, the LC develops a certain degree of nematic ordering
at densities or temperatures where the corresponding equilibrium system ($\dot\gamma=0$) is still globally isotropic.
Further, the director of the shear-induced nematic phase includes typically a non-zero angle with the direction of applied shear, commonly called the Leslie angle \cite{les68,rn01,rh99}. %Also, LCs show anisotropy in the viscosity which arises due to the possible alignments of the nematic director parallel to shear, gradient and vorticity directions \cite{mw46,shb14}. 
This term goes back to a phenomenological theory of the shear-induced dynamics of LCs provided by Leslie and Erikson \cite{les66,l99,ct07}, which successfully addresses various aspects of the rheological properties of LCs \cite{w81}. A more rigorous, statistical mechanics approach was developed by Doi {\em et al.} \cite{kd83} and Hess \cite{hes76}
who derived continuum equations for the nematic order parameter tensor from a (rotational) Fokker-Planck equation describing the non-equilibrium anisotropic fluid. Several aspects of these continuum theories, such as the resulting anisotropic viscosities, were later confirmed by computer simulations \cite{ss95,ss98,hsb90,sde93,se93,sec98}, where the LCs are modeled by GB ellipsoids \cite{gb81,mrcg91}. %Furthermore, Sarman et al., developed an algorithm to fix the director orientation during nonequilibrum MD simulations and demonstrated that the obtained viscosity coefficients for GB ellipsoids agrees with the equilibrium Green-Kubo relations \cite{sde93,se93,sec98}.  

Pure systems of dipolar spheres (which are often used as models of ferrofluids, i.e., suspensions of MNPs) also display shear thinning at large $\dot\gamma$ \cite{ich06,si15} and sufficiently large $\lambda$ (in the absence of an external magnetic field). Here, the shear thinning is related to the dipolar self-assembly into chains. At low densities where the equilibrium DSS fluid
forms isotropically distributed chains \cite{whm02,si13}, shear flow leads to a breaking of the chains. This is reflected by a reduction of the average size of the chains as function of $\dot\gamma$ \cite{si15,ich06}. Non-equilibrium Brownian dynamics and MD simulations suggest the DSS also show anisotropy in the viscosity under shear \cite{hsb90}. A theoretical description to explain the anisotropic viscosity of ferrofluids and the structural changes due to shear was proposed by Ilg {\em et al.}  \cite{ih03,kih03}. Also experiments indicate that the shear-induced changes in the viscosity of ferrofluids is strongly correlated with their structure, see, e.g., \cite{od03,so04}. In presence of an external magnetic field, the overall viscosity markedly increases due to chain formation in field direction \cite{po06,so08}. While most of these findings refer to dilute systems of dipolar spheres, there are also simulation studies at high densities \cite{mp02}. Here, one observes not only alignment of the chains along the shear direction, but also indications for shear-induced ferromagnetic ordering.

The above overview shows that the shear-induced behavior of LCs, on the one hand, and MNPs, on the other hand, is already quite well understood. This is not the case
for hybrid systems containing both components. From a general perspective, a LC matrix in its orientationally ordered state can be viewed as a viscoelastic medium. 
In a recent study, Ilg {\em et al.} \cite{ie18} have proposed a mesoscopic model to investigate the dynamics of (individual) MNPs in a viscoelastic medium, focusing 
on magnetic relaxation phenomena. The findings were found to be consistent with nanorheological experiments \cite{rrs14}. 
Still, it is clear that more particle-based investigations are needed to elucidate the behavior of magnetic hybrid systems with complex matrices under shear, especially
at larger dipolar coupling strengths.

From this perspective, we consider the model of GB and DSS particles considered in the present study as an archetypal example of a soft magnetic hybrid system. 
Our focus here is on the steady-state behavior in the absence of a magnetic field, particularly the shear stress and its interplay with chain formation and orientational ordering of the two components. At large $\lambda$, we find strongly nonlinear (non-Newtonian) behavior, while the more weakly coupled systems show a crossover from Newtonian to non-Newtonian behavior as functions of $\dot\gamma$.  Thus, the
flow properties of the mixture can indeed be tuned by varying the magnetic coupling.

The rest of the paper is organized as follows: in Sec.~\ref{sd} we present the details of the simulation and methods. Our observations regarding the behavior of the LC-DSS mixture under externally applied shear are discussed in Sec.~\ref{res}. Finally, in Sec.~\ref{con}, we conclude the paper with a short summary of the present work and an outlook towards future investigations.

\section{Simulation details}
\label{sd}
We consider a binary mixture of LC and DSS with a composition ratio 80:20 and perform non-equilibrium MD simulations using the LAMMPS package \cite{pl95,bpp09}. The LCs are modeled by ellipsoids which interact via a generalized Gay-Berne (GB) potential \cite{bpp09,bcm07,ee03}. We consider uniaxial LCs with an aspect ratio of 3:1, and the relative energy well depths for side-to-side interaction is considered to be five times stronger than the end-to-end interaction. The DSS interact via a combination of a soft sphere potential and dipolar interactions \cite{wp92,mdk06}. Finally, the interaction among the LC and DSS is modeled by a modified GB potential with the shape and energy parameters appropriate for spheres. We treat the long range dipolar interactions with the three dimensional Ewald sum \cite{mdk06,sk07,whm02}. A detailed description of interaction potentials and model parameters is available in Ref. \cite{gk19}.
 
The units of length and energy are set by the parameters of the GB potential, $\sigma_{0}$ and $\epsilon_{0}$, respectively as defined in Ref. \cite{gk19}. The parameters that characterize the structure and phase behavior of the mixture are the reduced temperature ${T}^{*} = k_{\textrm B}T/\epsilon_{0}$, the reduced number density $\rho^{*} = N \sigma_{0}^{3}/V$ (where $N$ and $V$ are the total number of particles and total volume, respectively), and the reduced dipole moment $\mu^{*} = \mu/\sqrt{\epsilon_{0}\sigma^{3}}$. The Newton's equations of motion for the force and torque acting on a particle (combined with a thermostat, see below) were integrated in the $NVT$ ensemble using a velocity Verlet algorithm and a reduced MD time step $\Delta t^{*} = \Delta t/\sqrt{m\sigma_{0}^{2}/\epsilon_{0}} = 0.002$. The simulations are performed at fixed $\rho^{*}$ and $T^{*}$ for various values of $\mu^{*}$. Specifically, $\mu^{*}$ ranges from 0.0 to 3.0, i.e., the values of dipolar coupling parameter $\lambda = \mu^{2}/k_{\textrm B}T\sigma_{0}^{3}$ ranges from 0.0 to 11.25. 
%%%%%%%%%%%%%%%%%%%%%%%%%%%%%%%%%%%%%%%%%%%%%%%%%%%%%%%%%%%%%%%%%%
\begin{figure}
 \centering
 \includegraphics[scale=0.65]{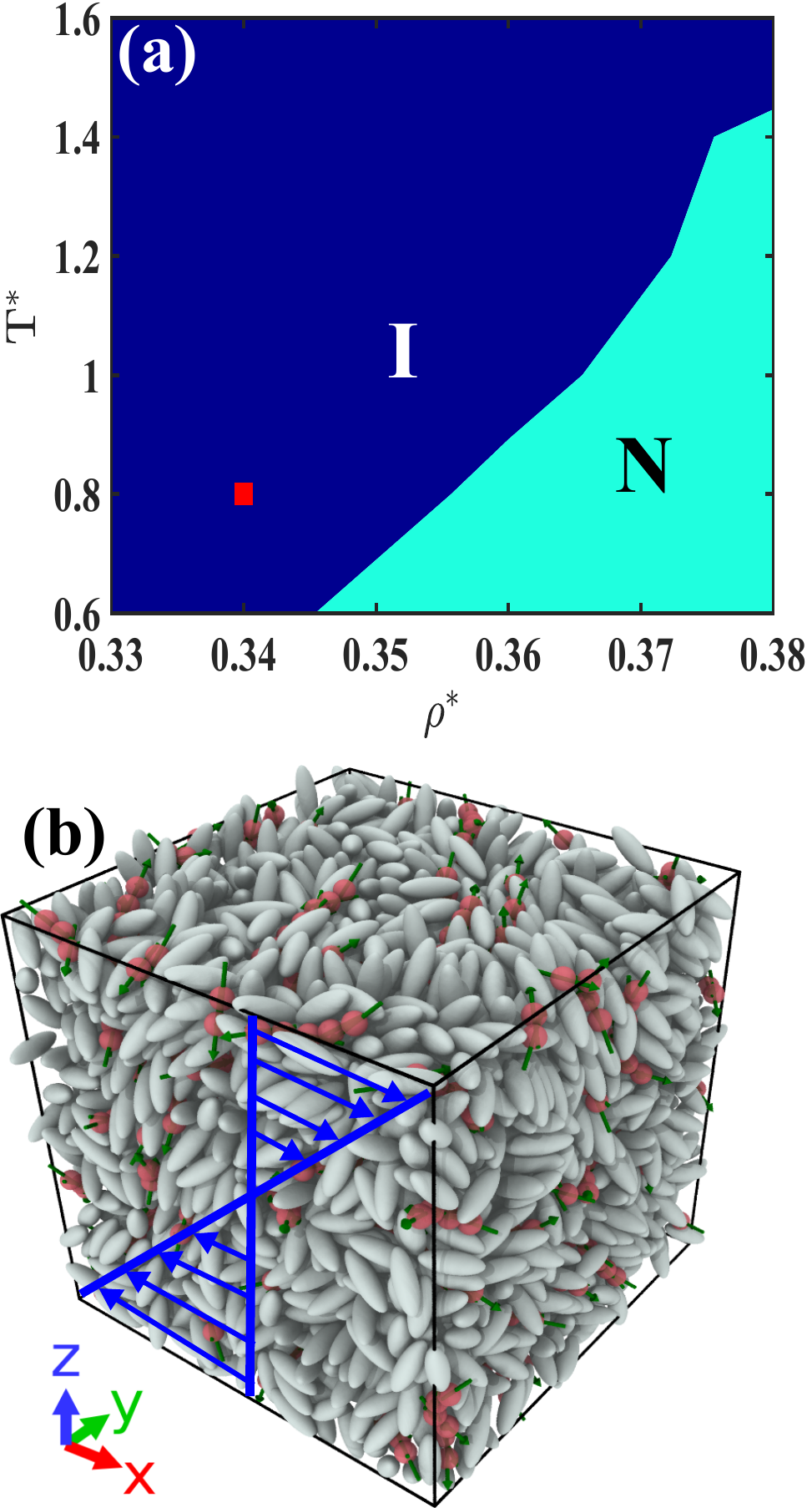}
 \caption{Shearing protocol for a LC-DSS mixture. (a) Tentative equilibrium phase diagram in the $\rho^{*}-{\rm T}^{*}$ plane obtained via MD simulations \cite{gk19}. The red square marks the quiescent state at $\rho^{*} = 0.34$, $T^{*} = 0.8$ and $\mu^{*} = 3.0$. (b) Snapshot of the mixture at the equilibrium on which a planar Couette flow is applied in the $x$-$z$ plane along the $x$-direction using Lees-Edwards periodic boundary conditions.}
 \label{snapeq}
\end{figure}
%%%%%%%%%%%%%%%%%%%%%%%%%%%%%%%%%%%%%%%%%%%%%%%%%%%%%%%%%%%%%%%%%%%

Our simulated system consists of 3200 LC ($N_{e}$) and 800 DSS particles ($N_{s}$). We start with a mixture equilibrated at $\rho^{*} = 0.34$ and $T^{*} = 0.8$. At these parameters, the mixture is in the isotropic phase, however, very close to the equilibrium I-N phase transition line \cite{gk19}.  The state point considered is marked by the red square in the equilibrium phase diagram of the LC-MNP mixture shown in Fig.~\ref{snapeq}(a) where the blue region represents the isotropic phase while the light green region shows the nematic phase \cite{gk19}. We obtain this tentative phase diagram by estimating the nematic order parameter at different state points.

We shear the mixture with constant shear rate using Lees-Edwards periodic boundary conditions (see the schematic diagram in Fig.~\ref{snapeq}(b) where the imposed velocity profile is shown by blue arrows). Specifically, we implement a planar Couette flow in the $x$-$z$ plane, with $x$ being the direction of flow, and $z$ and $y$ being the gradient and vorticity directions respectively. The range of dimensionless shear rates $\dot{\gamma}^{*} = \left( \sigma^{2}m/\epsilon\right)^{1/2}\dot{\gamma}$ is given by $10^{-3} - 10^{-1}$. To maintain the temperature we employ a Langevin thermostat acting in the gradient and vorticity directions \cite{em86}.

\section{Results}
\label{res}
\subsection{Stress-strain response}
\label{str}
 For the understanding of the rheology, the main quantity of interest is the shear stress, $\sigma_{xz}$, which we calculate via the Irving-Kirkwood expression \cite{at06,gs05},
\begin{eqnarray}
\label{str}
\sigma_{xz}(t) = \frac{1}{V}\left\langle\sum_{i}\left[m_{i}v_{i,x}v_{i,z} + \sum_{i>j} r_{ij,x}F_{ij,z}\right] \right\rangle.
\end{eqnarray}
In Eq.~(\ref{str}), $V$ is the total volume, $v_{i,(x,z)}$ are the $x$ and $z$ components of the velocity of  the ${\rm i}^{\rm th}$ particle and $r_{ij,x}, F_{ij,z}$ are the $x$-component of the distance vector and $z$-component of the force between particles $i$ and $j$, respectively.
 
\begin{figure*}
 \centering
 \includegraphics[scale=0.8]{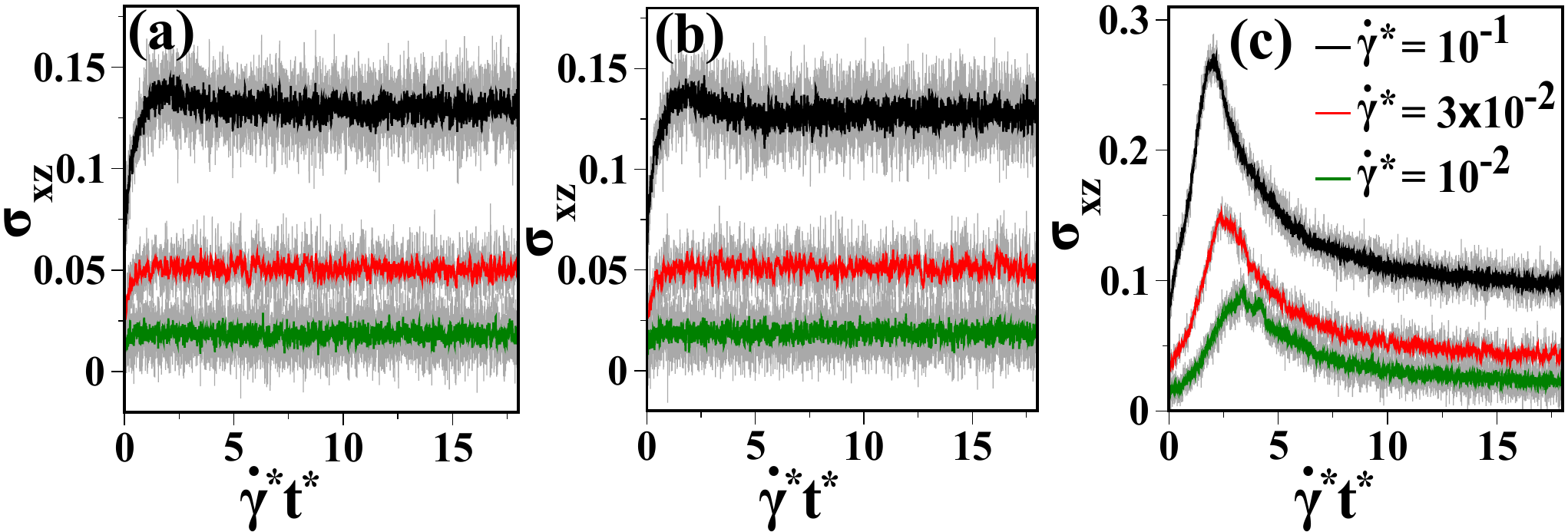}
 \caption{Evolution of $\sigma_{xz}$ as a function of strain $\dot{\gamma}^{*}t^{*}$ for the LC-DSS mixture at $\rho^{*} = 0.34$, $T^{*} = 0.8$ sheared with $\dot{\gamma}^{*} = 10^{-1}, 3\times 10^{-2}, 10^{-2}$ for (a) $\lambda = 0.0$, (b) $\lambda = 5.0$ and (c) $\lambda = 11.25$. Here, the shear rate is constant, therefore, the evolution of $\sigma_{xz}$ with strain $\dot{\gamma}^{*}t^{*}$ is equivalent to evolution with time. Gray curves in all plots represent data averaged over 100 samples. To further improve the statistics, a running average over 10 data points based on the gray curves has been performed.}
 \label{str_rp}
\end{figure*}

We start by investigating the time evolution of $\sigma_{xz}$ as a function of strain, $\dot{\gamma}^{*}t^{*}$, at fixed shear rate. In Fig.~\ref{str_rp}(a), (b) and (c), we present results for mixtures sheared with $\dot{\gamma}^{*} = 10^{-1}, 3\times 10^{-2}$ and $10^{-2}$ for $\lambda = 0.0, 5.0$ and $11.25$, respectively. 

At the largest shear rate considered ($\dot{\gamma}^{*} = 10^{-1}$), the stress first increases with strain, reaches a maximum and finally settles to a non-zero value. This ``overshoot" behavior, which appears for all values of $\lambda$, signals a complex behavior, it also appears in glassy systems and supercooled liquids \cite{vbb04,gch16}. As the shear rate is lowered, the height of the stress overshoot decreases and finally disappears for the mixtures with $\lambda = 0.0$ and $\lambda = 5.0$ (see Fig.~\ref{str_rp}(a),(b)). However, in the mixture with $\lambda = 11.25$ the stress overshoot persists even at the lowest shear rate considered. These observations already suggest that the degree of dipolar coupling plays a crucial role for the rheology of the mixture. We will come back to this point in the subsequent sections where we focus on the steady-state behavior.  The transient behavior, which appears interesting as well, will be discussed in more detail elsewhere \cite{gk20}.
%%%%%%%%%%%%%%%%%%%%%%%%%%%%%%%%%%%%%%%%%%%%%%%%%%%%%%%%%%%%%%%%%%
%%%%%%%%%%%%%%%%%%%%%%%%%%%%%%%%%%%%%%%%%%%%%%%%%%%%%%%%%%%%%%%%%%%
\subsection{Steady state flow curve}
\label{flow_curve}
\begin{figure}
 \centering
 \includegraphics[scale=0.65]{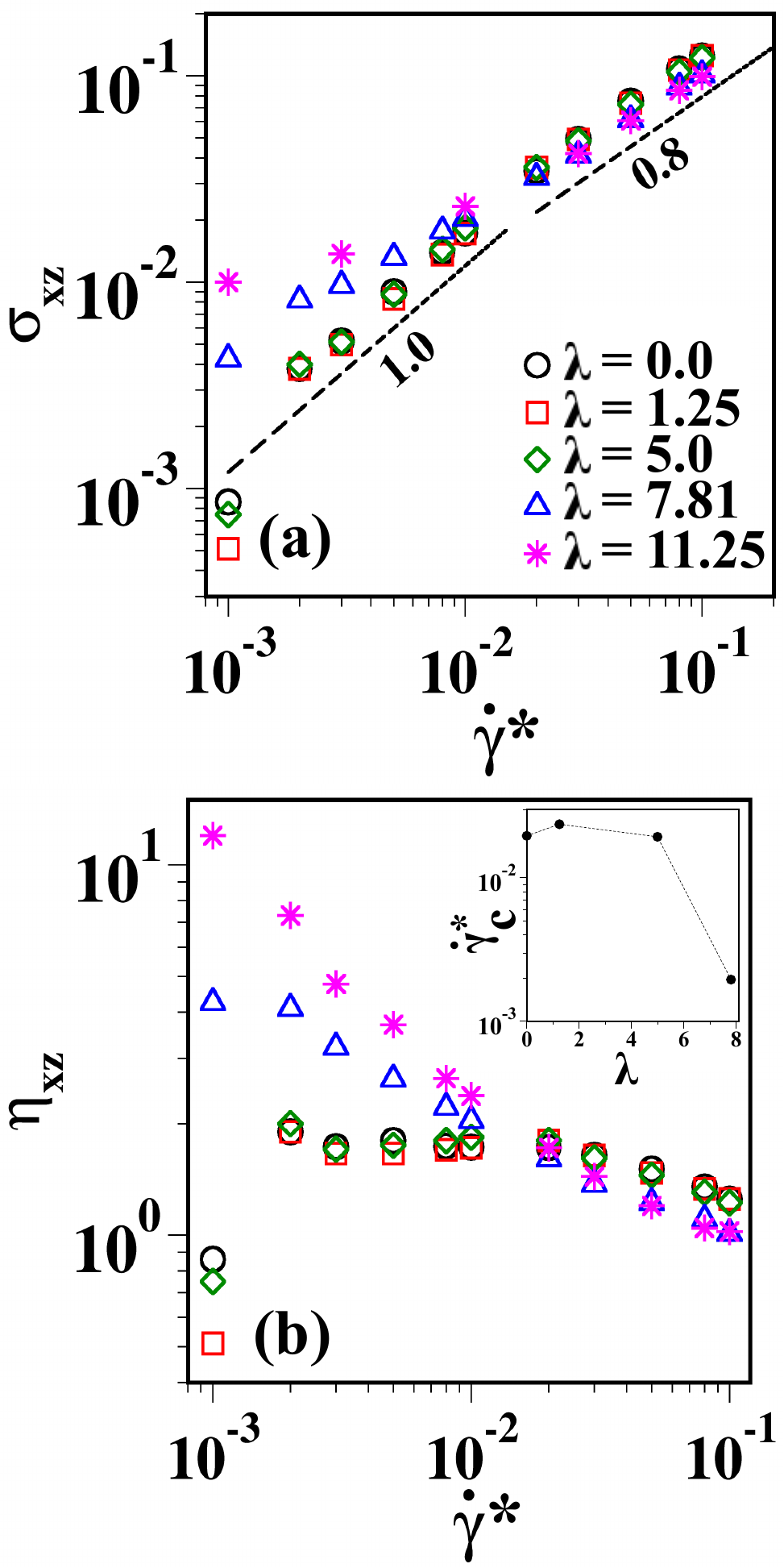}
 \caption{(a). The flow-curve for the LC-DSS mixture at $\rho^{*} = 0.34,~ T^{*} = 0.8$ and $\lambda = 0.0, 1.25, 5.0, 7.81, 11.25$. Dashed lines represent slopes 1.0 and 0.8. (b) Average viscosities $\eta_{xz}$, obtained from $\sigma_{xz}^{ss}$, as a function of $\dot{\gamma}^{*}$ for the same $\lambda$ values as in (a). Here, different symbols represent the same $\lambda$ value as in (a). The inset shows the variation of the critical shear rate $\dot{\gamma}_{c}^{*}$ at which a crossover from Newtonian to non-Newtonian behavior is observed, as function of $\lambda$.}
 \label{shearsnap}
\end{figure}

To proceed, we investigate the behavior of the steady-state stress, $\sigma_{xz}^{ss} = \lim_{t\to \infty} \sigma_{xz}$, as a function of the rate of the applied shear rate. Figure~\ref{shearsnap}(a) shows the resulting flow curve of the LC-DSS mixture for various values of $\lambda$. For all values of $\lambda$ except $\lambda = 11.25$, we observe a crossover from ``Newtonian" behavior, where $\sigma_{xz}^{ss}$ is linearly proportional to $\dot{\gamma}^{*}$, to ``non-Newtonian" behavior, where $\sigma_{xz}^{ss}$ varies in a power-law manner as a function of $\dot{\gamma}^{*}$. This transition occurs at a ``critical" shear rate, which is approximately given by $\dot{\gamma}^{*}_{c} = 0.02$. 

At shear rates $\dot{\gamma}^{*}\gtrsim \dot{\gamma}^{*}_{c}$ the stress varies as  $\sigma_{xz}^{ss}\propto \dot{\gamma}^{*n}$ for all values of $\lambda$, where the power-law exponent is $n\approx 0.8$. This exponent, also called flow index, characterizes the non-Newtonian behavior of the LC-DSS mixture. Specifically, the fact that for $\dot{\gamma}^{*}\gtrsim \dot{\gamma}^{*}_{c}$, $n$  is smaller than one, indicates a shear thinning. This behavior is also reflected by the average viscosity, defined as $\eta = \sigma_{xz}^{ss}/\dot{\gamma}^{*}$, which is plotted in Fig.~\ref{shearsnap}(b) as a function of $\dot{\gamma}^{*}$. For a wide range of complex fluids such as micorgels, foams, and emulsions, the exponent $n$ is considered to be a ``material parameter" that weakly depends on the temperature and density \cite{bdb17}. Here, we did not study this dependency systematically.

At large $\lambda (= 7.8125, 11.25)$ both the flow curve and the viscosity as function of $\dot{\gamma}^{*}$ reveal two distinct power-law regimes. This feature has also been observed in computer simulations of pure ferrofluids at high dipolar coupling strength \cite{si15}. Specifically, the steady-state stress grows slowly at low shear rates, before it crosses over to the power-law regime with slope $\sim 0.8$ at high shear rates. This behavior is more evident for $\lambda = 11.25$, where the non-Newtonian regime spans the whole range of shear rates considered in the simulations. 

%%%%%%%%%%%%%%%%%%%%%%%%%%%%%%%%%%%%%%%%%%%%%%%%%%%%%%%%%%%%%%%%%%
%%%%%%%%%%%%%%%%%%%%%%%%%%%%%%%%%%%%%%%%%%%%%%%%%%%%%%%%%%%%%%%%%%%

In contrast, at small values of $\lambda$ and low shear rates the stress grows linearly in $\dot{\gamma}^{*}$ and the viscosity remains constant, which is characteristic of Newtonian behavior. Shear-thinning behavior occurs only for $\dot{\gamma}^{*}> \dot{\gamma}^{*}_{c}$. We estimated critical shear rates $\dot{\gamma}_{c}^{*}$ for various $\lambda$, see the inset of Fig.~\ref{shearsnap}(b). The data reveal a sudden jump in $\dot{\gamma}_{c}^{*}$ when $\lambda$ increases beyond 5.0. Such a variation of $\dot{\gamma}_{c}^{*}$ with increasing $\lambda$ has already been observed for pure dipolar fluids, where it has been attributed to the chain forming tendency of the DSS at large $\lambda$ \cite{si15}. Similar behavior occurs in the present LC-DSS mixture, as will see below.

\subsection{Microstructure under shear} 
\label{structure}
The nonlinear features observed in the flow curves discussed in Sec.~\ref{flow_curve} already suggest profound structural changes when the LC-DSS mixture is exposed to shear. In the present section we focus, in particular, on the chain formation of magnetic particles. To this end we consider mixtures at $\lambda = 5.0$ and $\lambda = 11.25$, where the non-Newtonian behavior is particularly pronounced. Before we move ahead with the analysis, it should be noted that here in our LC-DSS mixture, the number density of the DSS is rather small $(\rho^{*}_{DSS} = 0.068)$. In pure DSS fluids, at such low densities, the DSS form chains with head-to-tail ordering of neighboring particles and the size of the chains depending on $\lambda$. These DSS chains are isotropically distributed, that is, there is no long-range orientational order \cite{si13,mdk06,sk07}. In our previous study of a LC-DSS mixture, we have shown that in the isotropic phase (such as one considered here at $\rho^{*} = 0.34$), the DSS form isotropically distributed chains of significant length if $\lambda \ge 6.0$ \cite{gk19}. Therefore, we expect that in the equilibrium, sizes of the DSS chains should be rather small for $\lambda = 5.0$ and relatively larger for $\lambda = 11.25$. We choose two shear rates, $\dot{\gamma}^{*} = 10^{-1}$ and $\dot{\gamma}^{*} = 5\times 10^{-3}$, for our analysis. For $\lambda = 5.0$, the shear rate $\dot{\gamma} = 10^{-1}$ falls into the non-Newtonian regime, while $\dot{\gamma}^{*} = 5\times 10^{-3}$ belongs to the Newtonian regime. For $\lambda = 11.25$, both the shear rates fall into the non-Newtonian regime, see Fig.~\ref{shearsnap}.

\begin{figure*}
 \centering
 \includegraphics[scale=0.8]{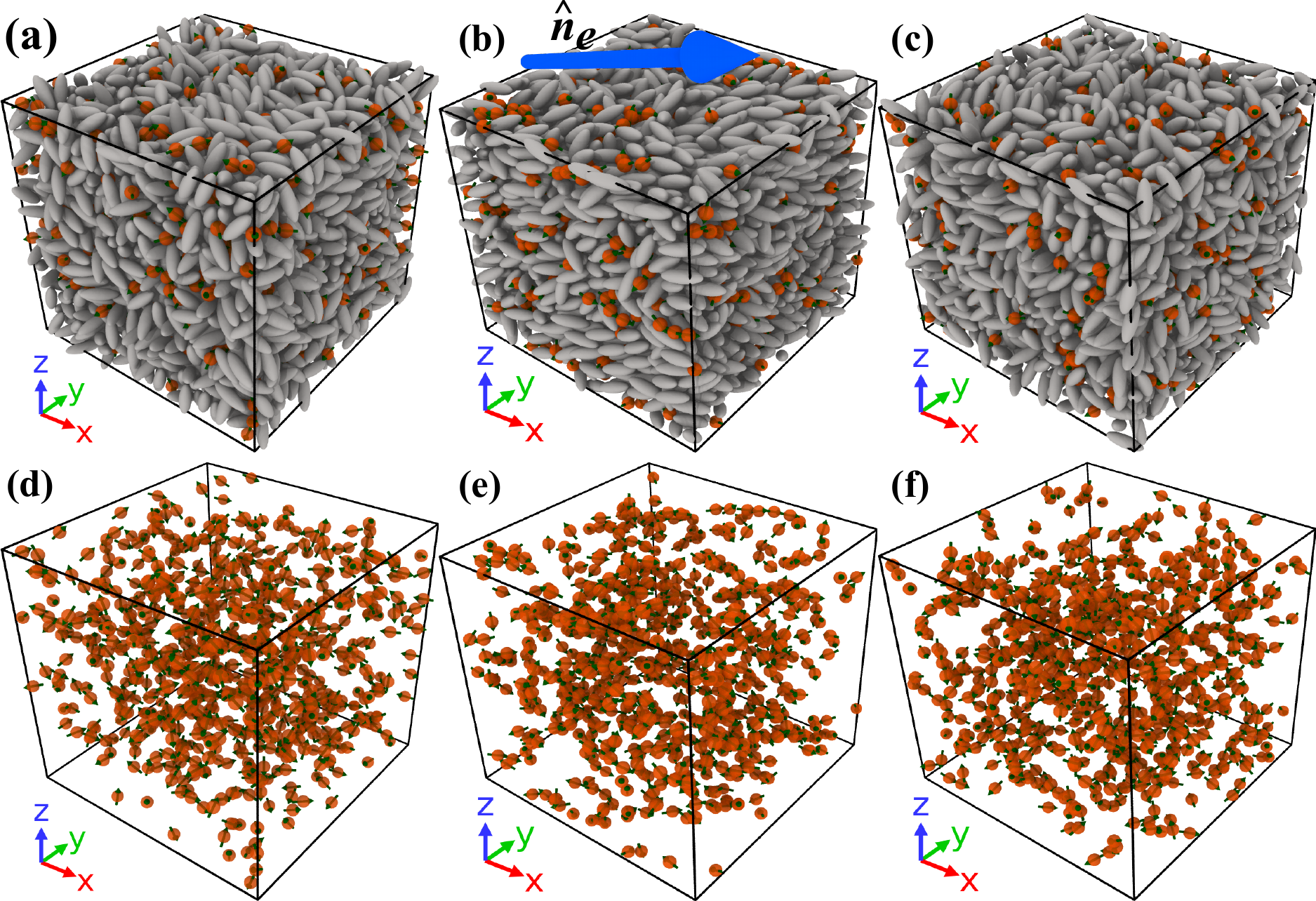}
 \caption{Snapshots of the LC-DSS mixture for $\lambda = 5.0$, $\rho^{*} = 0.34$ and $T^{*} = 0.8$ in equilibrium and under shear. (a) Initial state ($\dot{\gamma}^{*}t^{*} = 0.0$), which is same for all shear rates considered. (b) and (c) illustrate the structure in the steady state ($\dot{\gamma}^{*}t^{*} = 15.0$) for $\dot{\gamma}^{*} = 10^{-1}$ and $5\times 10^{-3}$, respectively. The snapshots (d), (e) and (f) show the DSS alone at the parameters corresponding to (b), (c), and (d) respectively. All the snapshots are prepared using software OVITO \cite{ov10}.}
 \label{shearsnapm2}
\end{figure*}

In Fig.~\ref{shearsnapm2}(a), we show a snapshot of the LC-DSS mixture for $\lambda = 5.0$ in the quiescent state at $\dot{\gamma}^{*}t^{*} = 0$. It is seen that the mixture consists of short DSS chains (shown in Fig.~\ref{shearsnapm2}(d)) and randomly oriented LCs. At the larger shear rate, $\dot{\gamma}^{*} = 10^{-1}$, LCs show a shear-induced alignment in the steady state, see Fig.~\ref{shearsnapm2}(b). This behavior is absent at the low shear rate $\dot{\gamma}^{*} = 5\times 10^{-3}$, see Fig.~\ref{shearsnapm2}(c). The DSS, plotted in Fig.~\ref{shearsnapm2}(e) and (f), show no alignment in the direction of the shear at any of the shear rates considered.
%%%%%%%%%%%%%%%%%%%%%%%%%%%%%%%%%%%%%%%%%%%%%%%%%%%%%%%%%%%%%%%%%%
%%%%%%%%%%%%%%%%%%%%%%%%%%%%%%%%%%%%%%%%%%%%%%%%%%%%%%%%%%%%%%%%%%%

\begin{figure*}
 \centering
 \includegraphics[scale=0.8]{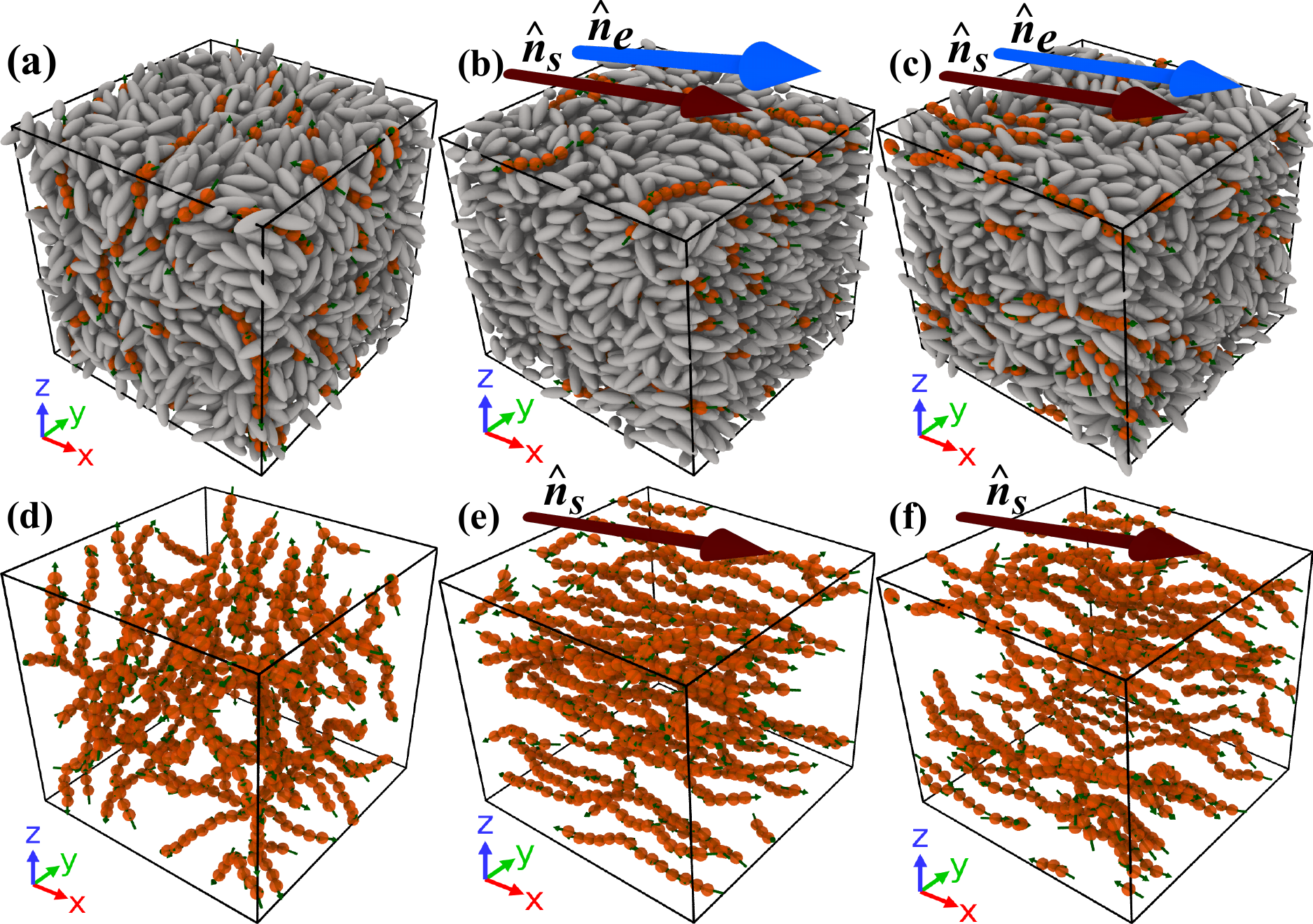}
 \caption{Snapshots of the LC-DSS mixture for $\lambda = 11.25$, $\rho^{*} = 0.34$ and $T^{*} = 0.8$ in equilibrium and under shear. (a) Initial state ($\dot{\gamma}^{*}t^{*} = 0.0$), which is same for all shear rates considered. (b) and (c) illustrate the structure in the steady state ($\dot{\gamma}^{*}t^{*} = 15.0$) for $\dot{\gamma}^{*} = 10^{-1}$ and $5\times 10^{-3}$, respectively. The snapshots (d), (e) and (f) show the DSS alone at the parameters corresponding to (b), (c), and (d) respectively.}
 \label{shearsnapm3r1}
\end{figure*}

In contrast to $\lambda = 5.0$, at large $\lambda (= 11.25)$ the equilibrium configuration is characterized by large DSS chains (see Fig.~\ref{shearsnapm3r1}(a) and (d)). In the presence of shear, these DSS chains align with the shear direction at both shear rates considered (see Fig.~\ref{shearsnapm3r1}(e) and (f)). The response of the LCs depends on the shear rate: while the LC system at $\dot{\gamma}^{*} = 10^{-1}$ (Fig.~\ref{shearsnapm3r1}(b)) displays clear alignment, the ordering at $\dot{\gamma}^{*} = 5\times 10^{-3}$ is less significant. We will quantify the degree of shear-induced nematic ordering in Sec.~\ref{orp}. Here we already note that the observed shear-induced ordering of the two species is consistent with previous simulation studies on pure LCs \cite{ss98,rhw08,gs05} and pure dipolar fluids \cite{si15,mp02}. The difference in the present case is that the two components strongly influence each other.
%%%%%%%%%%%%%%%%%%%%%%%%%%%%%%%%%%%%%%%%%%%%%%%%%%%%%%%%%%%%%%%%%%
%%%%%%%%%%%%%%%%%%%%%%%%%%%%%%%%%%%%%%%%%%%%%%%%%%%%%%%%%%%%%%%%%%
\subsection{Nematic order in the steady state}
\label{orp}
Along with the chain formation, a further interesting issue is the degree of shear-induced nematic ordering in both species of the LC-DSS mixture. To this end, we calculate the nematic order parameters $S_{e,s}$ via the largest eignvalues of the ordering tensors $\bm Q_{e,s}$ for the LCs ($e$) and the DSS ($s$) respectively. The instantaneous components of these tensors are given as
\begin{eqnarray}
Q_{\alpha\beta} = \left(1/N \right)\sum_{i=1}^{N}(1/2)\left(3{u}_{\alpha}^{i}{u}_{\beta}^{i}-\delta_{\alpha\beta}\right),
\end{eqnarray}
where $\alpha, \beta = x,y,z$. Further, for the $i^{th}$ particle, $u_{\alpha}^{i}$ is one of the components of the orientation unit vector $\hat{\bm u}$ in the case of LCs, while it is a component of the unit dipole vector $\hat{\bm \mu}$ in the case of DSS. The largest eigenvalue $\xi_{e,s}$ of the tensors $\bm Q_{e,s}$ characterizes the extent of the nematic ordering in the two components. The nematic order parameters $S_{e,s}$ are then obtained by taking the average of $\xi_{e,s}$ for many samples. Specifically, we have used 100 independent samples for calculation of order parameters and other quantities. Furthermore, the eigenvector corresponding to $\xi_{e,s}$ defines the nematic directors ${\hat{\bm n}}_{e,s}$ for the LCs and DSS respectively. The steady state values of $S_{e,s}$ are denoted by $S^{ss}_{e,s}$.

Furthermore, at large shear rates where nematic order is present, we analyze the orientation of $\hat{\bm n}_{e,s}$ with respect to the shear direction (i.e. with the $x$-axis in our case) by calculating the angle \cite{mp02},
\begin{eqnarray}
\label{theta}
\Theta_{e,s} = \langle atan(\mid \hat{\bm n}_{z,(e,s)}/\hat{\bm n}_{x,(e,s)}\mid) \rangle,
\end{eqnarray}
where $\hat{\bm n}_{z,(e,s)}$ and $\hat{\bm n}_{x,(e,s)}$ are the $z$ and $x$ components of the respective nematic directors. This angle defines the orientation of the projection of $\hat{\bm n}_{e,s}$ in the shear plane ($x$-$z$ plane) onto the shear direction ($x$-axis).

\begin{figure}
 \centering
 \includegraphics[scale=0.42]{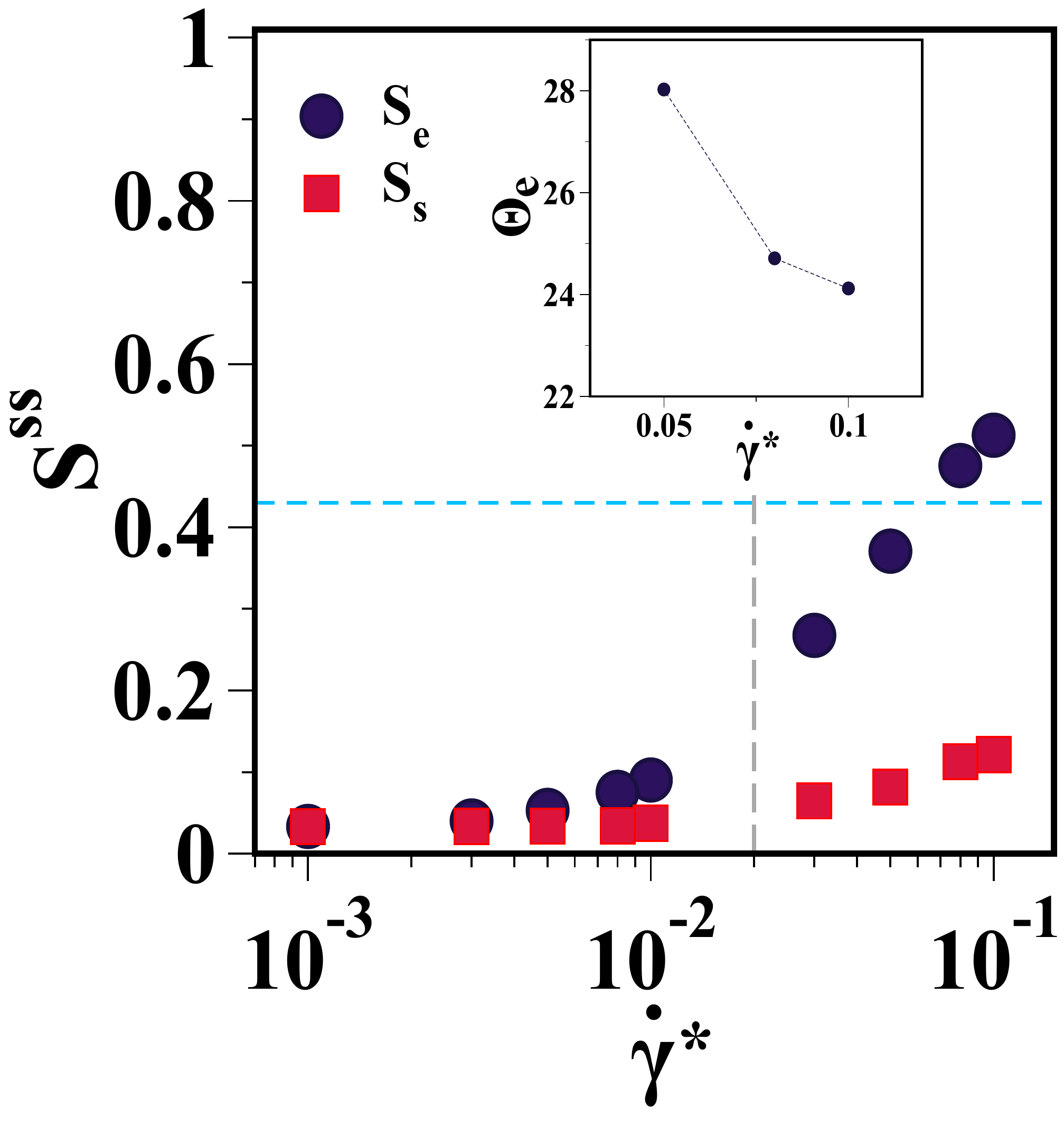}
 \caption{Variation of the nematic order parameter in the steady state for $\lambda = 5.0$ for the LC ($S_{e}$) and the DSS ($S_{s}$) as a function of $\dot{\gamma}^{*}$. The horizontal blue dashed line represents the critical value of the nematic order parameter at which I-N transition is observed and the vertical grey dashed line shows the $\dot{\gamma}_{c}^{*}$ for $\lambda = 5.0$. The inset shows the birefrigence angle $\Theta_{e}$ (in degrees) for the LCs at $\lambda = 5.0$. The three points in the inset correspond to $\dot{\gamma}^{*} = 10^{-1}, 8 \times 10^{-2}$ and $5 \times 10^{-2}$.}
 \label{orpshearm2}
\end{figure}

In Fig.~\ref{orpshearm2}, we plot the steady-state nematic order parameters $S^{ss}$ for the LC ($S_{e}$ black circles) and the DSS ($S_{s}$ red squares) for $\lambda = 5.0$ at different shear rates. We recall that at $\lambda = 5.0$, the flow curve reveals both, a Newtonian regime ($\dot{\gamma}^{*} < \dot{\gamma}^{*}_{c}$) and a non-Newtonian regime at high shear rates (see Fig.~\ref{shearsnap}). As seen from Fig.~\ref{orpshearm2}, the DSS do not develop any pronounced nematic ordering throughout the range of the shear rates considered. This is different for the LCs. In the Newtonian regime, the LCs do not show shear-induced ordering while at high shear rates, significant (para-)nematic ordering is observed. To this end we note that the value of $S_{e}$ related to the equilibrium I-N transition is 0.43 according to the Marier-Saupe theory \cite{gp03}. This value of $S_{e} (= 0.43)$ is represented by the horizontal light-blue dashed line in Fig.~\ref{orpshearm2}). One sees that this (equilibrium) value is approached and finally exceeded when the shear rate becomes larger than $\dot{\gamma}^{*}_{c}$ (indicated by the grey dashed line). More specifically, from the intersection of the function $S_{e}^{ss}(\dot{\gamma}^{*})$ and the horizontal blue line we estimate the shear rate of the shear-induced I-N transition as $\dot{\gamma}_{N}^{*}\approx 0.08$. We conclude that the non-Newtonian regime of the flow curve at $\lambda = 5.0$ is accompanied by orientational ordering of the LCs, but not the DSS.

The appearance of an isotropic-(para)nematic transition at high shear rates is consistent with previous simulation studies of GB ellipsoids \cite{msn01}, attractive colloidal rods \cite{rhw08,rhjp08} and soft repulsive long ellipsoids \cite{gs05}. 

The inset in Fig.~\ref{orpshearm2} shows the angle $\Theta_{e}$ between the director of the LCs and the $x$- axis (see Eq.~\ref{theta}). Note that, since significant nematic ordering is present only at large shear rates, the relevant points in the inset are those corresponding to $\dot{\gamma}^{*} = 8\times 10^{-2}$ and $10^{-1}$. From these, we observe that the nematic director for LCs does not entirely align with the shear direction, but makes a small angle of approximately $24^{\circ}$. This observation is consistent with previous studies on pure Gay-Berne LCs under shear \cite{wqz07,msn01}.

%%%%%%%%%%%%%%%%%%%%%%%%%%%%%%%%%%%%%%%%%%%%%%%%%%%%%%%%%%%%%%%%%%
%%%%%%%%%%%%%%%%%%%%%%%%%%%%%%%%%%%%%%%%%%%%%%%%%%%%%%%%%%%%%%%%%%%

\begin{figure}
 \centering
 \includegraphics[scale=0.65]{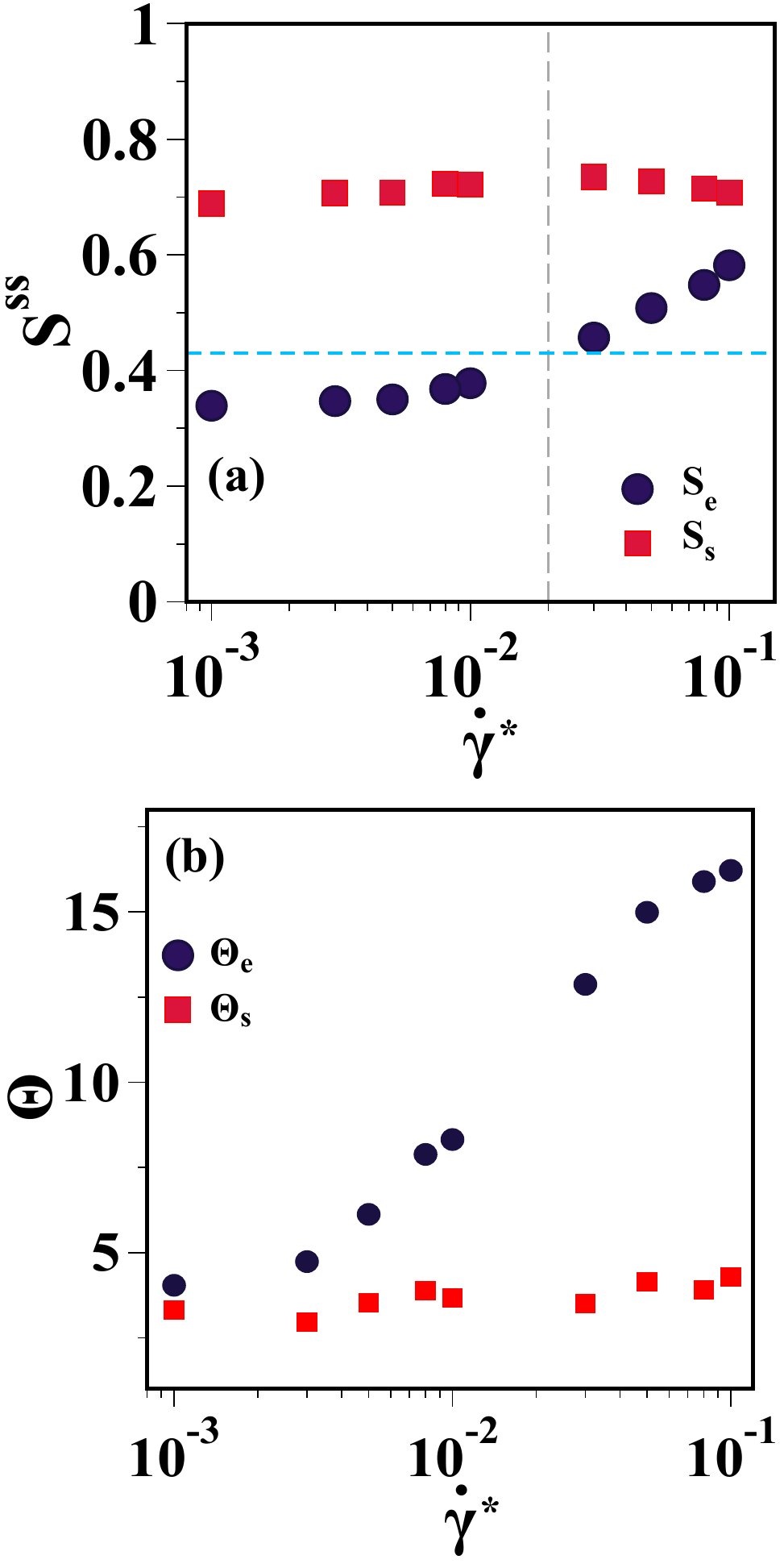}
 \caption{(a) Variation of the nematic order parameters for the LC ($S_{e}$) and the DSS ($S_{s}$) in the steady state for $\lambda = 11.25$ as a function of $\dot{\gamma}^{*}$. The horizontal blue dashed line represents the value of the nematic order parameter corresponding to the I-N transition. The vertical grey dashed line indicates $\dot{\gamma_{c}^{*}}$ for $\lambda = 11.25$. (b) Birefringence angle $\Theta$ (in degrees) as a function of $\dot{\gamma}^{*}$ for the LC and the DSS at $\lambda = 11.25$}
 \label{orpshearm3}
\end{figure}

The situation at large $\lambda$, shown in Fig.~\ref{orpshearm3}(a), is different. Here, the LC matrix exhibits a significant degree of ordering already at the lowest shear rate, $\dot{\gamma}^{*} = 10^{-3}$. Moreover, the nematic order parameter of the DSS is even larger. We understand these properties, which are in marked contrast to those observed at $\lambda = 5.0$, as a consequence of the pronounced chain formation of the DSS at $\lambda = 11.25$, see Fig.~\ref{shearsnapm3r1}(e) and (f).

Due to the strong correlation within the chains, these form rather stiff and long objects which align in the shear flow. In fact, they align essentially along the shear ($x$-) direction, as seen from the small values of the angle $\Theta_{s}$ plotted in Fig.~\ref{orpshearm3}(b). The alignment of the chains, in turn, enhances the alignment of the non-magnetic LC particles, leading to relatively large values of $S^{ss}_{e}$. 

Upon increase of the shear rate, the nematic order parameter of the DSS remains essentially constant in the range of the shear rates considered. For pure dipolar fluids in the shear flow,  it has been observed that the nematic order parameter decreases at high shear rates due to breaking of the chains \cite{si15,mp02,kih03}. In the present study, we do not observe such a behavior as the maximum shear rate is restricted to $\dot{\gamma}^{*} = 0.1$.

Finally, we consider the birefrengence angles $\Theta_{s}$, defined in Eq.~(\ref{theta}) and plotted in Fig.~\ref{orpshearm3}(b). For all considered shear rates the director of the DSS makes an angle of about $3^{\circ}$ from the $x$-axis, indicating that the DSS chains almost completely align along the shear direction in the steady state.

In contrast, the nematic director of the LCs, makes an angle $\Theta_{e}$ of about $16^{\circ}$ at large shear rates with the shear direction. However, $\Theta_{e}$ decreases as the shear rate is lowered. At the same time, the degree of nematic ordering decreases as well, see Fig.~\ref{orpshearm3}(a). We suspect that at these low shear rates, the LCs locally align along the DSS chains, that is, along the $x$-direction, which may lead to the observed reduction in $\Theta_{e}$.

%%%%%%%%%%%%%%%%%%%%%%%%%%%%%%%%%%%%%%%%%%%%%%%%%%%%%%%%%%%%%%%%%%
%%%%%%%%%%%%%%%%%%%%%%%%%%%%%%%%%%%%%%%%%%%%%%%%%%%%%%%%%%%%%%%%%%%
\subsection{Ferromagnetic ordering}
\label{fmo}
Given the pronounced nematic ordering of the dipolar chains at large values of $\lambda$ (see Fig. 7a)), it is an obvious question whether this nematic ordering is accompanied by {\em ferromagnetic} ordering, i.e., parallel orientation of the chains. To this end we consider the polar order parameter 
\begin{eqnarray}
\langle P_{1}\rangle = \bigg\langle\frac{1}{N} \left| \sum_{i = 1}^{N} {\hat{\bm\mu}}_{i}\cdot {\hat{\bm n}}_{s}\right|\bigg\rangle,
\end{eqnarray}
where ${\hat{\bm\mu}_{i}}$ is the unit dipole vector of the $i^{th}$-particle, and ${\hat{\bm n}}_{s}$ is the nematic director of the DSS (the angular brackets denote the average over different samples). Perfect ferromagnetic order corresponds to $\langle P_{1}\rangle=1$.

As a background information we note that pure DSS fluids do indeed
display ferromagnetic ordering (in three dimensions) at large dipolar coupling strengths ($\lambda\gtrsim 6.67$) and large densities ($\rho\sigma^3\gtrsim 0.7$) \cite{wp92,mdk06}. Within a mean-field picture \cite{zw95,agp95,agp97}, the ferromagnetic ordering can be explained via the non-zero average field generated by the neighbors around a dipolar sphere, provided that the boundary conditions are appropriate (i.e., conducting). The relevance of the boundary conditions is a consequence of the long-range character of the dipolar interactions. The mean field is proportional to the density \cite{zw95} and dipolar coupling strength, which allows for ferromagnetic ordering in dense systems and sufficiently low temperatures. From a structural point of view, ferromagnetic ordering is related to a specific configuration of neighboring dipolar chains, where particles are shifted by half a particle diameter \cite{wpa92,wl93}. In this situation, the interaction between two chains is indeed attractive \cite{wpa92}.

In the present mixture, the number density of the dipolar component is much smaller than in the cases mentioned above, $\rho\sigma^3\approx 0.068$. Thus, at least in equilibrium, one would not expect ferromagnetic ordering even at the largest $\lambda$ considered, and this is confirmed by simulations \cite{pk15}. The question then is whether such an ordering can be induced by the shear flow. Indeed, shear-induced ferromagnetic ordering has been observed for pure, dense DSS fluids at coupling strength $\lambda \sim 4.63$, where the corresponding unsheared system is still isotropic \cite{mp02,lmp02}. The density considered in Ref.~\cite{mp02,lmp02}, however, was $\rho^{*}=0.8$, that is, far beyond that of the present system.

\begin{figure}
 \centering
 \includegraphics[scale=0.65]{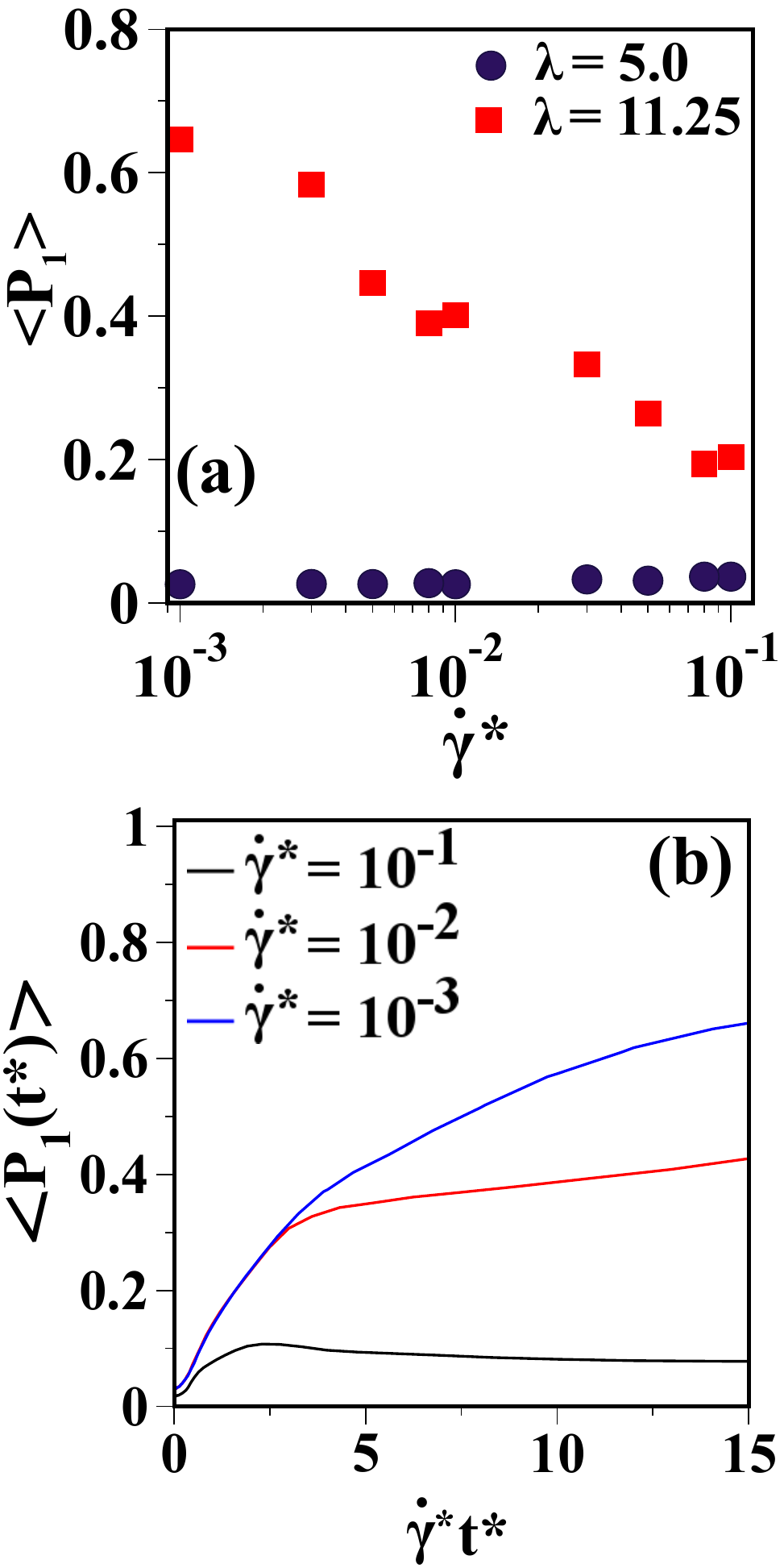}
 \caption{(a) Ferromagnetic order parameter $\langle P_{1}\rangle$ in the steady state as a function of $\dot{\gamma}^{*}$ for $\lambda = 5.0$ and $11.25$. (b) The evolution of $\langle P_{1}\rangle$ with strain for $\dot{\gamma}^{*} = 10^{-1}, 10^{-2}$ and $10^{-3}$.}
 \label{magshear}
\end{figure}

Numerical results for $\langle P_{1}\rangle$ in the steady state are plotted in Fig.~\ref{magshear}(a) as function of the shear rate. For $\lambda=5.0$, the values of the order parameter are negligible throughout the regime considered. However, for $\lambda=11.25$ and the lowest shear rate, $\dot{\gamma}^{*}=10^{-3}$, we observe a quite large value of $\langle P_{1}\rangle\lesssim 0.7$, reflecting a significant degree of ferromagnetic order. This large value decreases upon increase of $\dot{\gamma}^{*}$.
%%%%%%%%%%%%%%%%%%%%%%%%%%%%%%%%%%%%%%%%%%%%%%%%%%%%%%%%%%%%%%%%%%
%%%%%%%%%%%%%%%%%%%%%%%%%%%%%%%%%%%%%%%%%%%%%%%%%%%%%%%%%%%%%%%%%%%

Given the low density, the large order parameter at $\dot{\gamma}^{*}=10^{-3}$ is indeed quite surprising. To check that the ordering is indeed induced by shear we have also investigated $\langle P_{1}\rangle$ as function of strain $\dot{\gamma}^{*} t^{*}$, see Fig.~\ref{magshear}(b). The data clearly reveal the absence of ferromagnetic order at zero shear. The curve for $\dot{\gamma}^{*}=10^{-3}$ also indicates that the ferromagnetic order is gradually "built up" when the strain is increased. Further, the values at the largest strain ($\dot{\gamma}^{*} t^{*}=15$) reflect what is already seen in Fig.~\ref{magshear}(a), namely, that the degree of ferromagnetic order in the steady state decreases with the applied shear rate.

There remains the question as to {\em why} the system at low shear rate orders at all, despite of the low density. From a mean-field perspective, one can argue as follows: Due to the nematic ordering of the LC host matrix (which, in turn, aligns the dipolar chains), the possible directions of ferromagnetic order of the dipolar component are strongly restricted as compared to the three-dimensional, pure dipolar fluid. Indeed, in the pure case, the director can assume any direction on the unit sphere, whereas in presence of the nematic matrix, the direction of magnetization is essentially restricted to two directions, along the shear and opposite to shear. This corresponds to a change from "XYZ symmetry" to "Ising symmetry", i.e., a profound reduction of degeneracy. Mean-field theory for a pure dipolar system would then predict a strong reduction of the critical density beyond which ferromagnetic order can occur. Indeed, this critical density is given by $\rho^{*}_c = 9/(4\pi\lambda)$ for conventional, three-dimensional dipoles, while it is only $\rho^{*}_c = 3/(4\pi\lambda)$ for an "Ising" dipolar system which can order only "up" or "down" \cite{agp95,agp97}. Clearly, this line of argumentation assumes an effective equilibrium picture of the system at low shear rates. Still, it describes one important mechanism which could promote ordering at lower densities (even though the theory is known to be strongly incorrect in terms of quantitative predictions).

\begin{figure}
 \centering
 \includegraphics[scale=0.45]{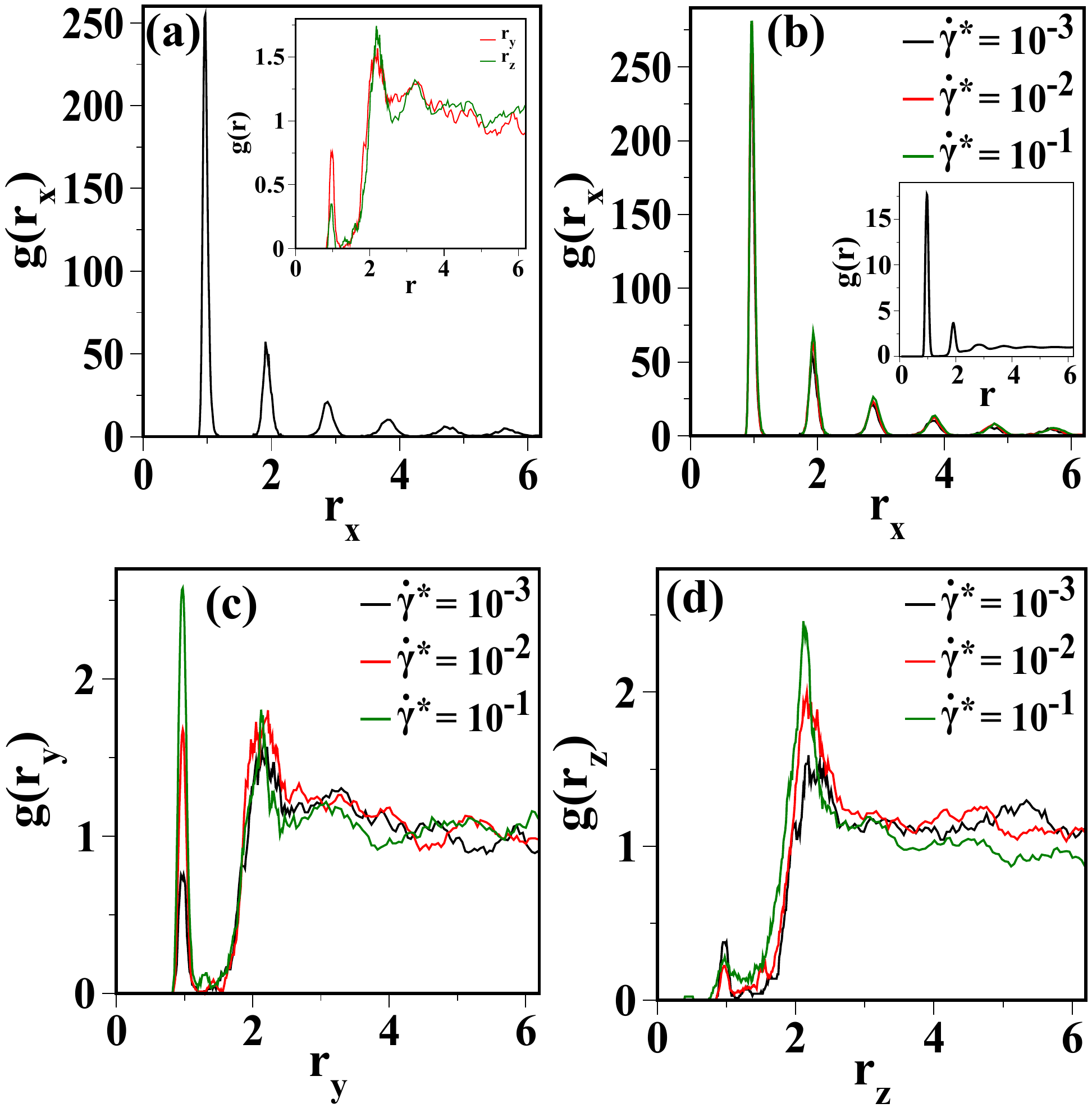}
 \caption{Pair correlation functions between the DSS at $\lambda = 11.25$. (a) $g(r)$ as a function of distance in the $x$-direction for the shear rate $\dot{\gamma}^{*} = 10^{-3}$. The inset shows the correlations in $y$ and $z$-directions for the same shear rate as in the main figure. (b), (c) and (d) correlation function for three different shear rates $\dot{\gamma}^{*} = 10^{-3}, 10^{-2}$ and $10^{-1}$ in the $x$, $y$ and $z$-directions respectively. The inset in (b) shows the radial distribution function $g(r)$ as a function of $r$ in equilibrium.}
 \label{mom}
\end{figure}

A further piece of understanding emerges when we consider the structure around a dipolar chain and its dependence on $\dot{\gamma}^{*}$. To this end we have plotted in Fig.~\ref{mom} various pair correlation functions involving the DSS alone.
%
%%%%%%%%%%%%%%%%%%%%%%%%%%%%%%%%%%%%%%%%%%%%%%%%%%%%%%%%%%%%%%%%%%
%%%%%%%%%%%%%%%%%%%%%%%%%%%%%%%%%%%%%%%%%%%%%%%%%%%%%%%%%%%%%%%%%%%

Specifically, we have calculated separately the DSS pair correlation functions in directions of shear ($x$-direction), the shear gradient ($z$-direction), and vorticity ($y$-direction) (for a definition of these correlations, see Ref. \cite{wl93}). Fig.~\ref{mom}(a) shows all three correlation function at the lowest shear rate, where the order parameter $\langle P_{1}\rangle$ is largest. We observe pronounced, nearly solid-like correlations in shear direction, which corresponds to the direction of the chains. These correlations are, in fact, much stronger than those in the equilibrium system (see inset of Fig.~\ref{mom}(b)), indicating that the shear flow, via the ordering of the LC host matrix, strongly promotes the chain formation. The correlations in the other two directions (see inset of Fig.~\ref{mom}(a)) are much less pronounced (and quite similar). Interestingly, the second peak of these lateral correlation function is much higher than the first one. This strongly differs from the equilibrium situation, where the first peak exceeds by far the other ones. This feature reflects that shear strongly promotes correlations {\em between} neighboring chains; in fact, they appear to be more close than in equilibrium. We recall that in the ferromagnetic phase of dense, pure DSS systems, chains are close and are shifted by half a particle diameter, because only then their interaction is attractive \cite{wpa92}. Therefore, a small chain-chain distance is an important ingredient into the development of ferromagnetic order. 

Finally, Figs.~\ref{mom} (b)-(d) illustrate the dependency of the correlation functions on the shear. It is seen that the height of the first peak in lateral directions somewhat increases with increasing shear rate, indicating that shear enhances the correlation. This is somehow surprising in view of the fact that $\langle P_{2}\rangle$ decreases with $\dot{\gamma}^{*}$ (see Fig.~\ref{magshear}(a)). One possible explanation could be a breakage of chains which is not directly reflected by the correlation functions plotted in Fig.~\ref{mom}. This point remains to be considered in a future study.

\section{Summary and outlook}
Based on non-equlibrium MD simulations, we have shown that the strength of the dipolar coupling plays a crucial role for the rheological properties of LC-DSS mixtures. In particular, it changes the ``critical'' shear rate at which the flow curve transforms from Newtonian to non-Newtonian behavior in the steady state.

Further, we find that shear modifies the spatial structure within the mixture and induces an alignment in both components. For low values of $\lambda$, the DSS chains are relatively short and do not align with the shear direction. In contrast, for large values of $\lambda$, we observe long DSS chains aligning with the shear direction, yielding a nematic ordering of the chains. The degree of this shear-induced nematic ordering is almost independent of the rate of applied shear, and it is significantly higher than the nematic ordering of the LC matrix.

The LC matrix shows a shear-induced nematic transition at high shear rates for all values of $\lambda$. This transition occurs at a shear rate which shifts to lower values as $\lambda$ increases. In particular, at the largest $\lambda$ considered, nematic ordering of the LC matrix already appears at very low shear rates, although the equilibrium system is isotropic. This is due to the fact that the long DSS chains formed at large values of $\lambda$, align with the shear direction already at low shear rates and thereby induce a nematic ordering in the LCs. Similar behavior is also observed in isotropic (unsheared) LC-DSS mixtures in the presence of an external magnetic field, where the alignment of the DSS chains in the field direction leads to a nematic ordering in the LCs \cite{mespkk16}. Here, the ordering effect of a magnetic field is replaced by shear flow.

Finally, we have found indications that, at very large $\lambda$, the sheared mixture displays not only nematic ordering, but also ferromagnetic order of the DSS chains. This is a striking observation in view of the small density of the DSS, which we attempted to explain by mean-field arguments. It seems clear, however, that further investigations are necessary to unravel the underlying structural mechanism.

In conclusion, our results on the steady-state rheological properties of the LC-DSS mixture provide interesting suggestions to design soft hybrid materials with tunable flow properties. To further understand the full rheology of these systems, investigations of the time evolution (that is, the transient structure) of the microstructure under shear and on non-equilibrium flow patterns are in order. Walk in these directions is on the way. A further interesting research direction concerns the rheology of LC-DSS mixtures with non-spherical particles. First results have been reported in Ref. \cite{ssk19}.
\label{con}
\section{Acknowledgments}
We gratefully acknowledge funding support from the Deutsche Forschungsgemainschaft (DFG) via the priority program SPP 1681.
%%%%END OF MAIN TEXT%%%
%
%%The \balance command can be used to balance the columns on the final page if desired. It should be placed anywhere within the first column of the last page.
%
%%If notes are included in your references you can change the title from 'References' to 'Notes and references' using the following command:
%%\renewcommand\refname{Notes and references}

%%%REFERENCES%%%
\bibliography{ref} %You need to replace "rsc" on this line with the name of your .bib file
\bibliographystyle{rsc}

%%%%%%%%%%%%%%%%%%%%%%%%%%%%%%%%%%%%%%%%%%%%%%%%%%%%%%%%%%%%%%%%%%%
\end{document}